\documentclass[a4paper,11pt]{article}
\usepackage{jinstpub} 
\usepackage{lineno}
\usepackage{orcidlink}
\usepackage{upgreek,xspace}
\usepackage{subfigure}
\usepackage{caption}
\usepackage{subcaption}
\def\textmu{\ensuremath\upmu}

\title{New candidate polymeric wavelength shifters for noble liquid detectors}

\author[a,1]{M.~Kuźniak\orcidlink{0000-0001-9632-9115}\note{Corresponding author.}}
\author[a]{S.~Choudhary}
\author[b]{S.~Pawłowski}
\author[a]{A.\,F.\,V.~Cortez}
\author[b]{M.~Kaczorowski}
\author[b]{M.~Kumosiński}
\author[b]{A.~Abramowicz}
\author[c]{T.~\L ęcki}
\author[a]{G.~Nieradka}
\author[a]{T.~Sworobowicz}
\author[b]{D.~Jamanek}

\affiliation[a]{AstroCeNT, Nicolaus Copernicus Astronomical Center of the Polish Academy of Sciences, Rektorska 4, 00-614 Warsaw, Poland}
\affiliation[b]{Łukasiewicz Research Network – Industrial Chemistry Institute, Rydygiera 8, 01-793 Warsaw, Poland}
\affiliation[c]{Biological and Chemical Research Centre, Faculty of Chemistry, University of Warsaw, \.Zwirki i Wigury 101, 02-089 Warsaw, Poland}

\emailAdd{mkuzniak@camk.edu.pl}

\abstract{Polymeric wavelength shifters are of particular interest for large liquid argon detectors. 
Inspired by the success of polyethylene naphthalate (PEN), other new polymers exhibiting a similar type of excimer fluorescence were investigated.
We report on the preliminary results of the first cryogenic wavelength shifting test of a solution-cast film of PVN, poly(2-vinyl naphthalene). Significant brittleness was identified as a factor potentially limiting the use of PVN. However, clear signs of wavelength shifting were observed, with the overall efficiency and time response comparable to those of PEN. 
}


\begin{document}
\maketitle
\flushbottom

\section{Introduction}
Liquid argon detectors for dark matter searches, long-baseline neutrino experiments and neutrinoless double-beta decay searches, rely on wavelength shifters (WLS) with visible blue emission for detection of vacuum ultraviolet (VUV) argon scintillation light at 128~nm~\cite{instruments5010004}. 
Given 100-10,000~m$^2$ of surface area of WLS, anticipated in the next generation detectors, industrial grade large format polymeric polyethylene naphthalate foils (PEN) have been proposed~\cite{Kuźniak2019} and adapted to specific detector applications~\cite{Boulay2021}.

Increasing the wavelength shifting efficiency (WLSE) remains an important goal, which directly impacts the physics sensitivity. In parallel to R\&D on enhancing PEN  with customized synthesis~\cite{Kuźniak_2024}, we have commenced a broader survey of candidate polymeric WLS materials.

\section{Poly(vinyl naphthalene)}
In the case of PEN, the dominant emission mechanism is the fluorescence transition of an excimer state formed between two adjacent naphthalene groups.
\begin{figure}[htb]
    \centering
    \subfigure[polyethylene naphthalate]{\includegraphics[width=0.31\linewidth]{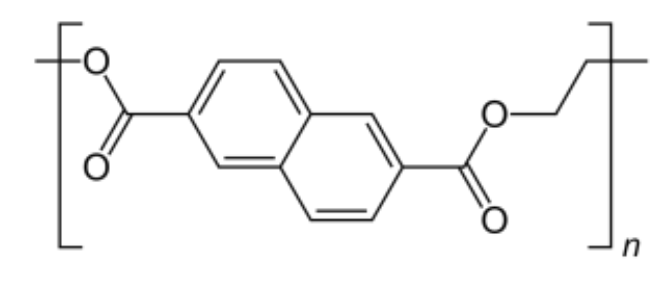}}
    \subfigure[poly(1-vinyl naphthalene)]{\centering\includegraphics[width=0.25\linewidth]{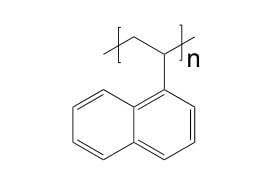}}\subfigure[poly(2-vinyl naphthalene)]{\includegraphics[width=0.28\linewidth]{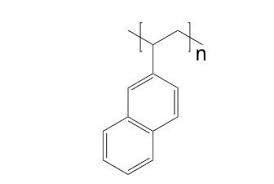}\label{a}}
    \caption{Structures of polymers exhibiting excimer emission from adjacent naphthalene units: PEN, P1VN and P2VN.}
    \label{fig:schematic}
\end{figure}

Assuming that the overall efficiency of the process is correlated with the number density of such pairs of adjacent units in the material, two other promising and relatively simple polymers can be indicated, see Fig.~\ref{fig:schematic}. PVN maximizes the relative contribution of naphthalene groups, reducing the non-luminescent polymeric skeleton to a bare minimum.

Indeed, excimer emission with a broad peak at 410~nm, resembling that of PEN and well matched to the spectral sensitivity of commonly used photon detectors such as photomultipliers or silicon photomultipliers, has been reported in the literature for both PVN species. David et al.~\cite{DAVID19721019} reported excimer fluorescence  emitted only by P1VN films, with approximately a factor of 2~increase in emission intensity between the room temperature and 77~K, under a 300~nm excitation wavelength. Frank~\&~Harrah~\cite{frank} also observed monomer emission in P2VN, peaked at 340~nm, however only for thin films (5~\textmu m). They attributed the disappearance of monomer emission of thicker PVN samples to self-absorption effects. For P2VN, a smaller increase in excimer emission intensity by a factor of~1.4 was reported with the temperature decreasing from room temperature to 87~K. Webber~\cite{Webber} reported the time constants for the P2VN monomer and excimer emission as 7.4~ns and 34.5~ns, respectively.

All the above makes PVN a suitable wavelength shifter candidate, worth testing under conditions more representative of a liquid argon detector. To this end, three samples of P2VN powder with varying chain lengths were procured: $M_n=48,000$ (product ID 10928) and 914000 (product ID 11013) from Polymer Source, Inc., and $M_w=175,000$ from Merck. 

\section{Room temperature fluorescence}
\begin{figure}
    \centering
        \includegraphics[width=0.65\linewidth]{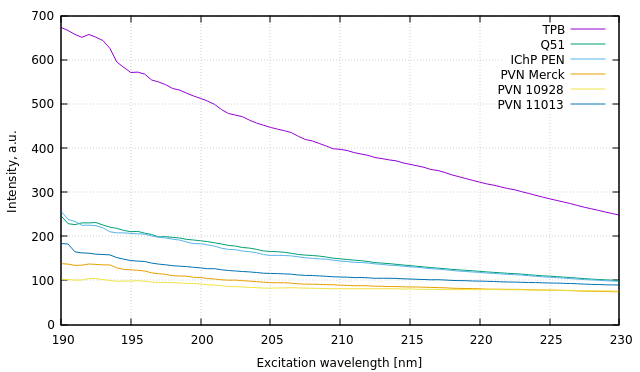}
    \caption{Fluorescence intensity of the three tested P2VN pressed powder samples, compared to a commercial grade PEN Teonex~Q51 foil, custom synthesised PEN~\cite{Kuźniak_2024} chunk, and a tetraphenyl butadiene coated specular reflector reference, as a function of excitation wavelength and at room temperature.}
    \label{fig:pvn}
\end{figure}
As an initial test, pressed powder samples of all the polymers have been checked for fluorescence using a Shimadzu UV-3600 spectrophotometer equipped with a 15~cm diameter integrating sphere
accessory (LISR-3100), as in Refs.~\cite{Kuźniak2019,Boulay2021,Kuźniak_2024}.

Clear hints of P2VN fluorescence were visible, with the brightest emission from the longest chain sample, although its intensity was somewhat lower than that of PEN samples, see Fig.~\ref{fig:pvn}.

\section{Foil sample production}
\begin{figure}
    \centering
        \includegraphics[width=0.3\linewidth]{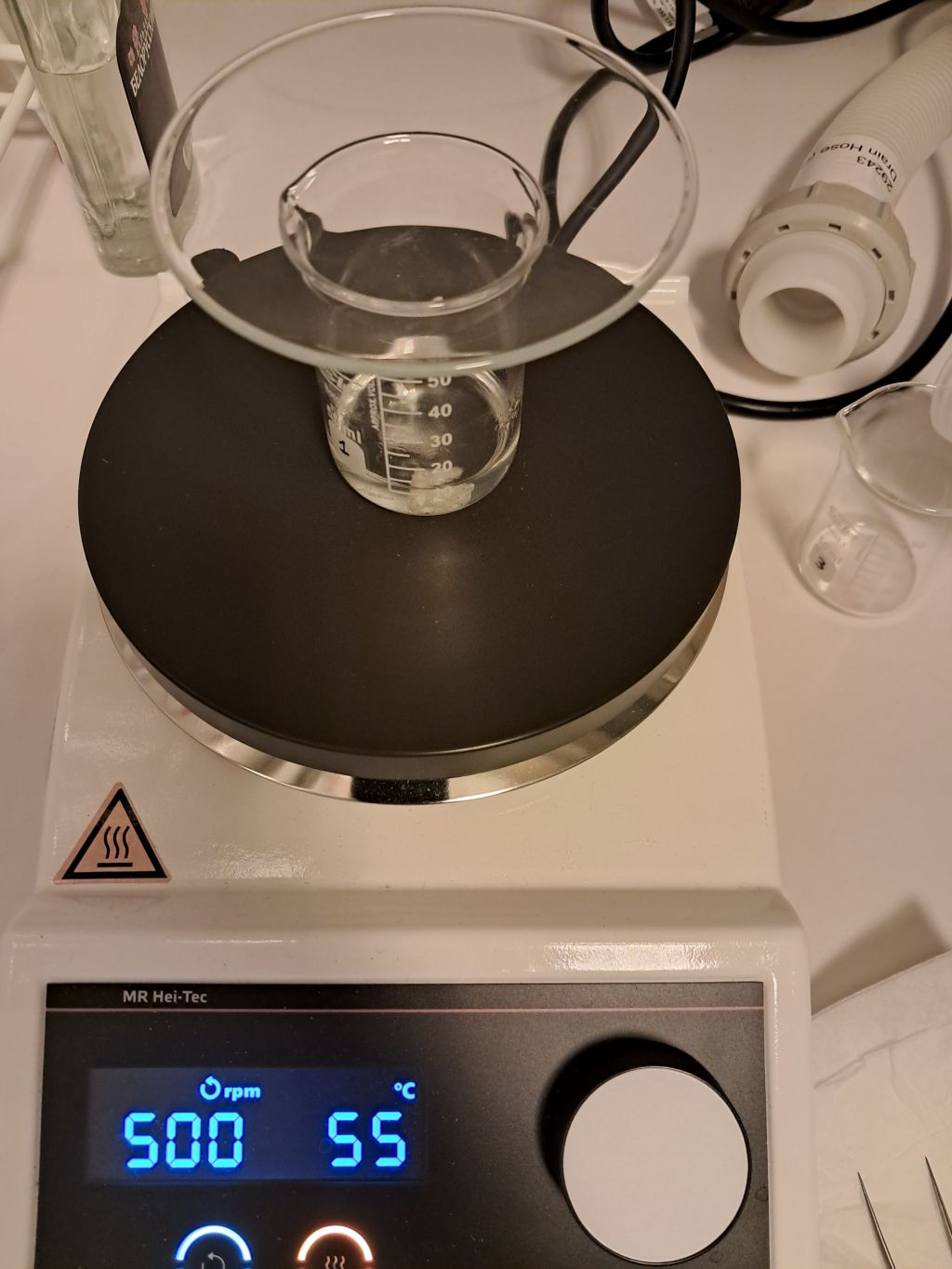}~\includegraphics[width=0.3\linewidth]{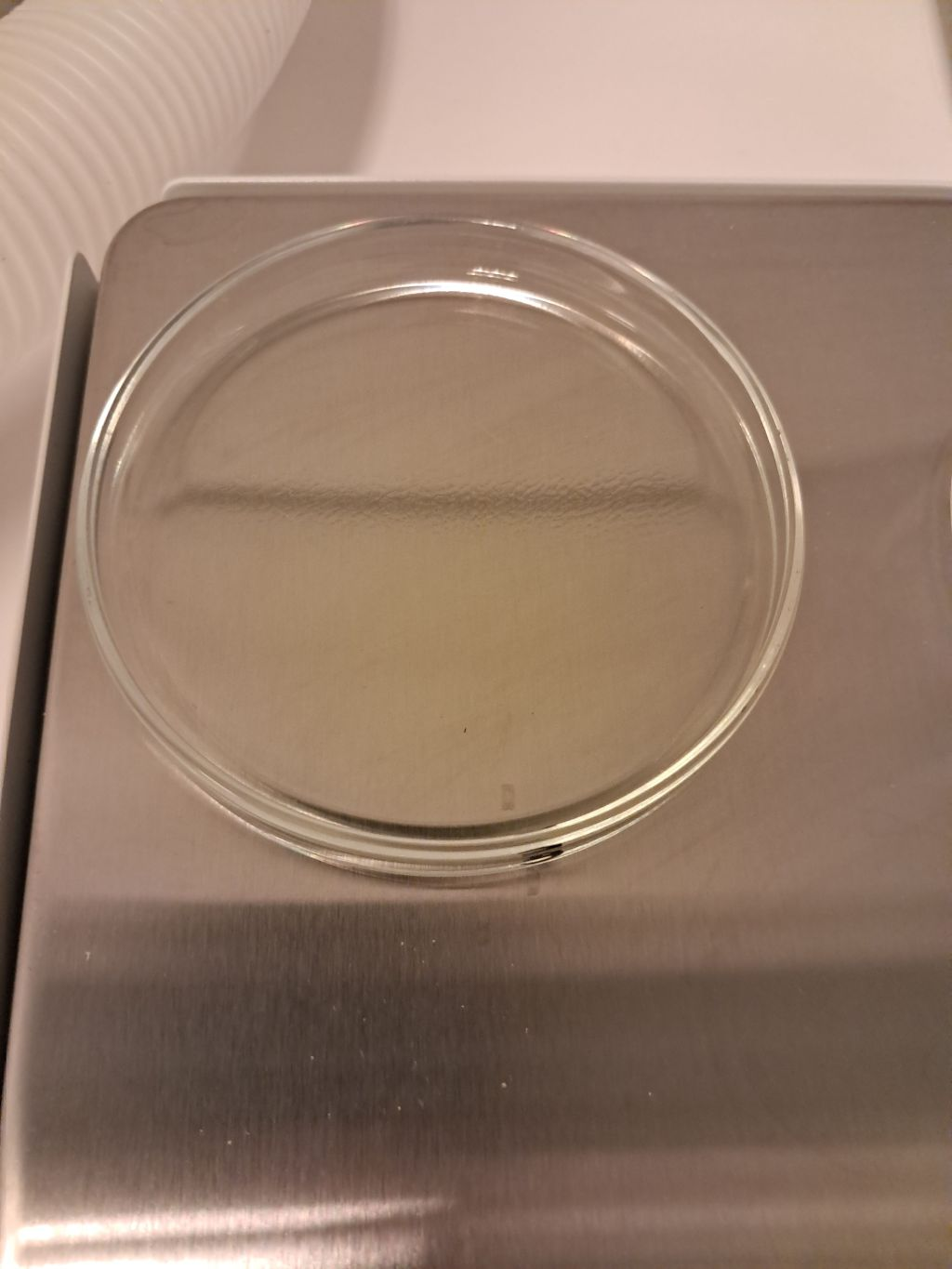}
        \includegraphics[width=0.3\linewidth]{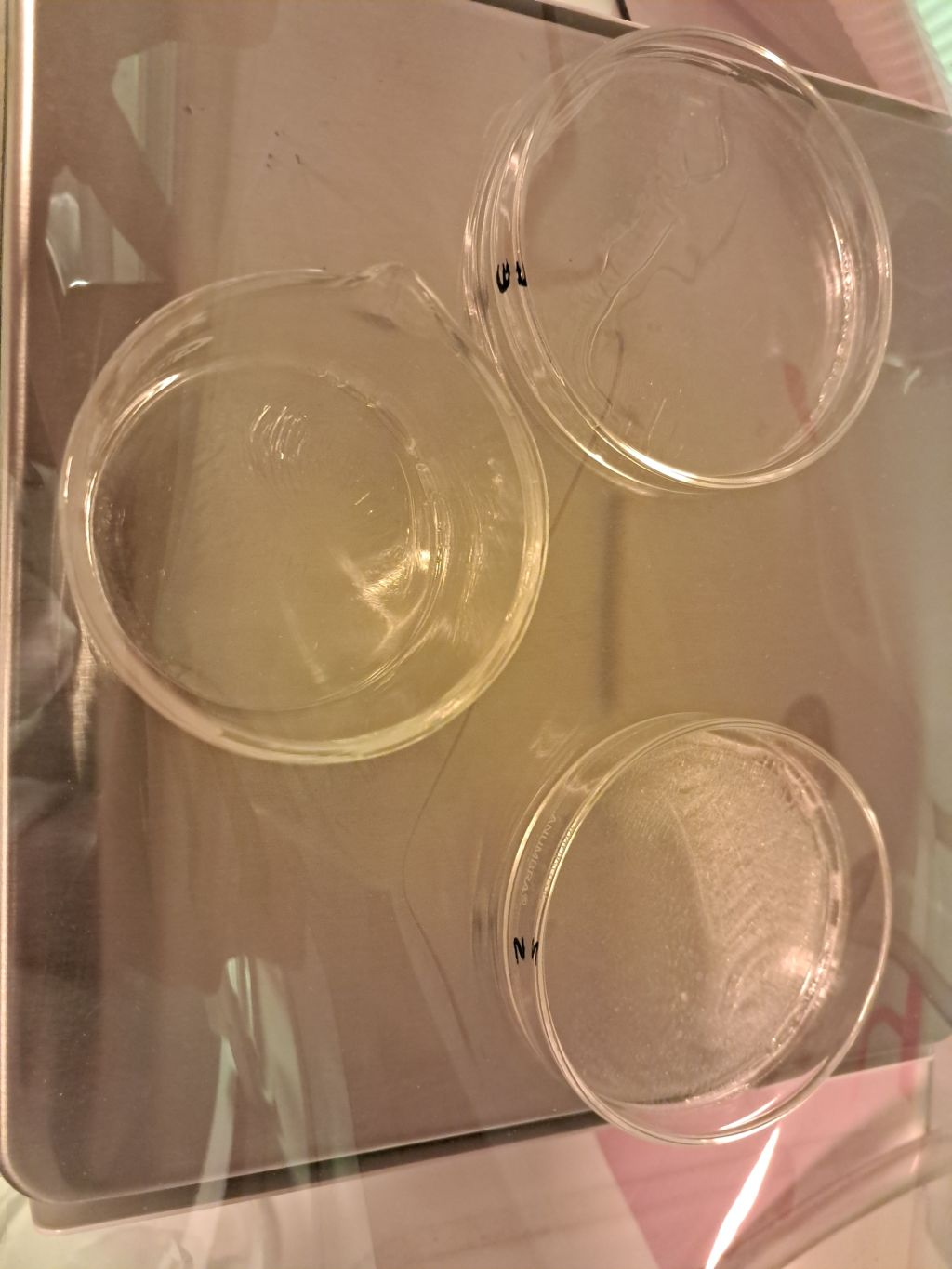}~\includegraphics[width=0.3\linewidth]{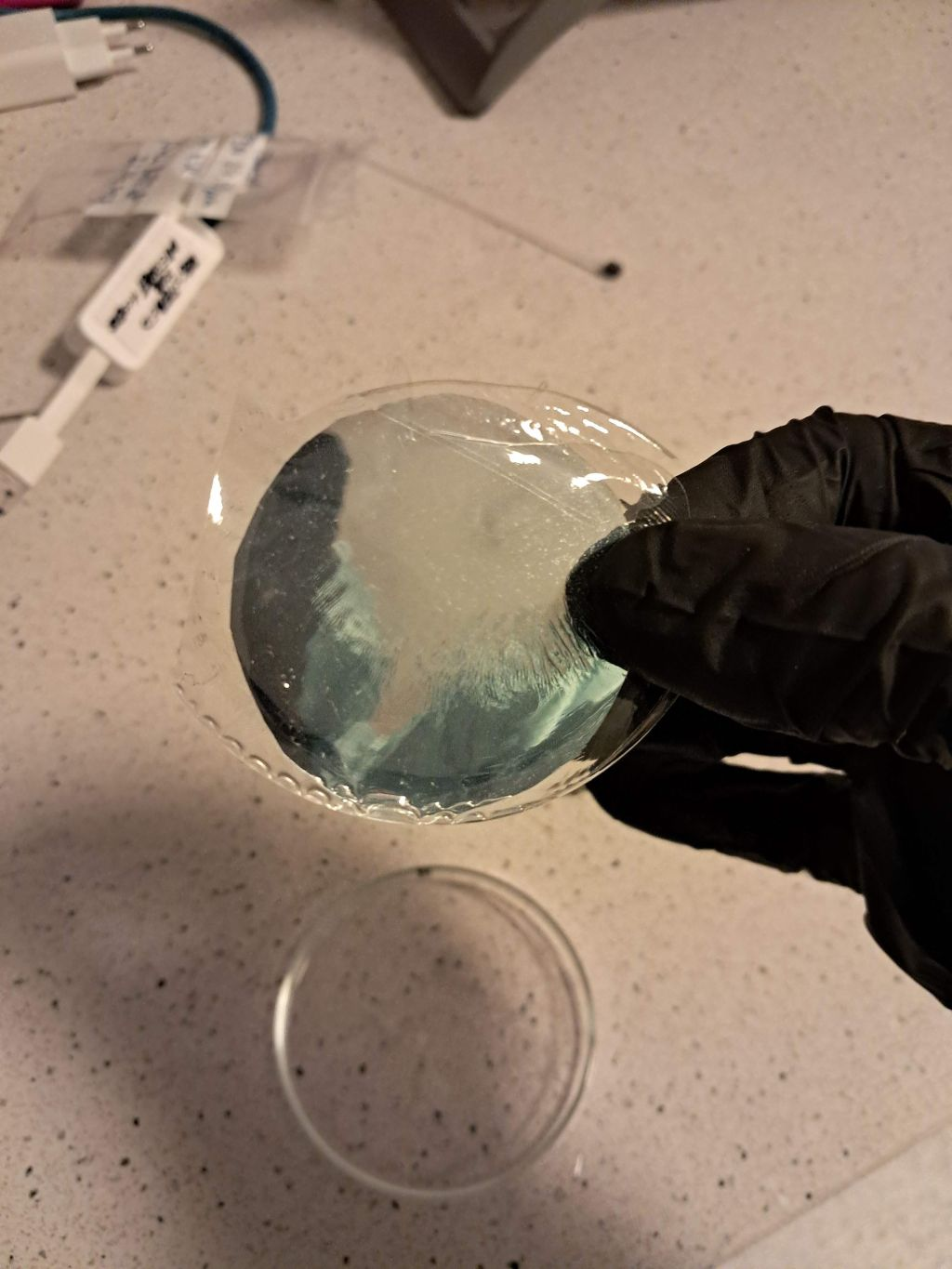}
    \caption{Casting of P2VN foil samples. (Top left) Dissolving in toluene on a hot plate with a magnetic stirrer. (Top right) Cast wet film on a Petri dish. (Bottom left) Foil samples after drying. (Bottom right) The P2VN sample attached to a specular reflector before the installation in a cryostat sample holder.}
    \label{fig:casting}
\end{figure}

A standalone self-supporting foil sample was then required for a cryogenic fluorescence measurement in ArGSet~\cite{choudhary2024cryogenic}. Unlike PEN, PVN dissolves easily in many common organic solvents, making it relatively simple to cast foil samples from solution. P2VN samples were dissolved in toluene and cast into foils, aiming for an approximate dry thickness of 100~\textmu m , then dried overnight, as shown in Fig.~\ref{fig:casting}.

Due to fractures densely covering the entire surface of the short chain length sample, it could not be used as a self-supporting foil for the measurement. The long chain length sample formed a uniform film, however was too brittle to handle. In the end, only the medium chain length sample was used for the measurements, despite its significant brittleness.

\section{Cryogenic wavelength shifting efficiency test}
ArGSet is a small cryostat filled with gaseous argon. Argon scintillation at 128~nm is induced by alpha particles depositing their energy in the volume between the WLS sample and two SiPMs, facing the sample. SiPMs are blind to 128~nm light and detect only visible light wavelength-shifted in the sample.

\begin{figure}
    \centering
    \includegraphics[width=\linewidth]{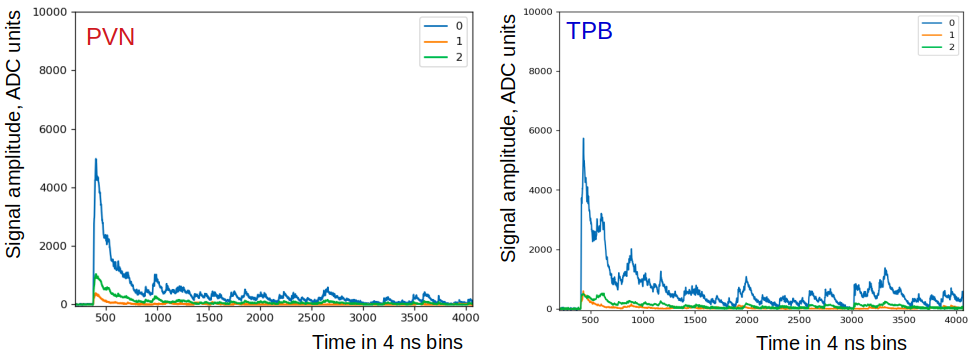}
        \includegraphics[width=0.7\linewidth]{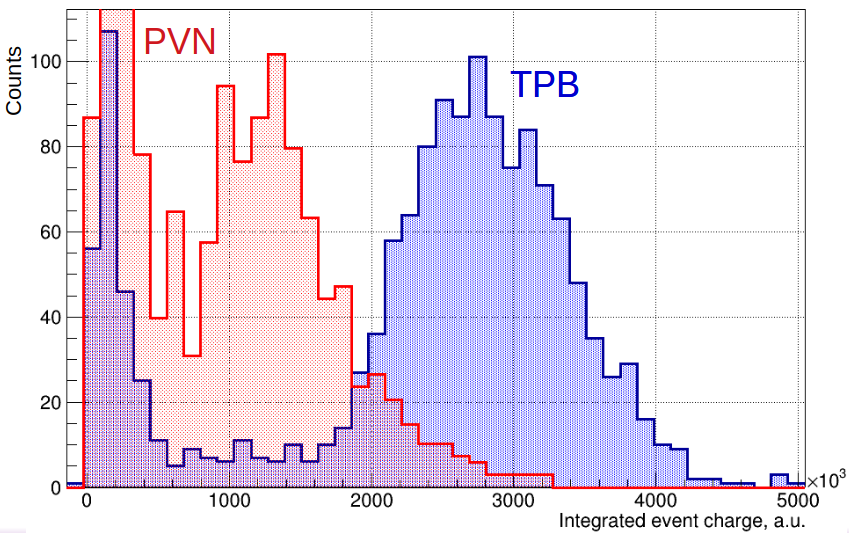}
    \caption{(Top) SiPM waveforms registered for typical events for PVN and TPB, for each of "1" and "2" SiPMs and in the analogue sum channel "0" (with additional amplification). (Bottom) Integrated charge spectra for P2VN sample (red) and tetraphenyl butadiene (blue).}
    \label{fig:pvntpb}
\end{figure}
Two runs were performed, one with the medium chain length P2VN sample, and followed directly by a run with a tetraphenyl-butadiene-coated reflector, used as a reference sample. Clear waveforms, with features typical of argon gas scintillation were observed for both samples. The spectra of integrated waveform charge from both runs are shown in Fig.~\ref{fig:pvntpb}, with clear peaks attributed to the alpha induced events. The location of these peaks is proportional to the emission intensity of the wavelength shifter and, assuming otherwise
similar conditions, can be used for a relative comparison of wavelength shifting efficiency of the samples.

By comparing the locations of both peaks, the WLSE of P2VN is in the range of 40-50\% of TPB -- an indicative result only, as the systematic uncertainty of the setup is currently being re-evaluated. However, this suggests that
P2VN is comparable in wavelength shifting efficiency to PEN. 

\section{Room temperature scintillation test}
To check for the scintillation properties of P2VN, the medium chain length foils was placed directly on SiPMs of the ArGSet, at room temperature, with an $^{241}$Am alpha source placed on top. A few minutes later, a similar test was conducted with a 125~\textmu m PEN foil sample (from Goodfellow), at the same SiPM voltage. PEN is well known for its excellent scintillation properties~\cite{Nakamura_2011}, however significant differences in scintillation yield between grades have been previously reported, and no absolute calibration of the reference PEN sample was available at the time of the measurements. The integrated intensity of the peak events registered with PVN is 73\% of those registered in PEN foil, which is an indication of the relative scintillation yield of both materials.     

\section{Summary and outlook}
A family of poly(vinyl naphthalene) polymers was identified as a potentially suitable class of wavelength shifters for cryogenic experiments. Preliminary results indicate the P2VN film wavelength shifting efficiency under vacuum ultraviolet excitation and in cryogenic conditions of 40-50\% of TPB, which is similar in value to PEN, and with time constants of a similar order. Under exposure to alpha particles
at room temperature, P2VN also exhibits scintillation properties, with the yield of approx. 73\%~of Teonex. However, the solution-cast films of P2VN used for tests were very brittle, which would prevent their use in a manner similar to PEN (i.e.,
as large-format foils in liquid argon). A measurement with proper analysis of experimental uncertainties for both P2VN and P1VN is planned next. 

\acknowledgments
This work has been supported by the Polish National Science Centre (UMO-2022/47/B/ST2/02015), from the EU’s Horizon 2020 research and innovation programme under grant agreement No 952480 (DarkWave), and from the International Research Agenda Programme AstroCeNT (MAB/2018/7) funded by the Foundation for Polish Science from the European Regional Development Fund~(ERDF). We are grateful to Prof. Magdalena Skompska for access to the spectrophotometer, which was purchased by CNBCh (University of Warsaw) from the project co-financed by EU from the ERDF.

\bibliographystyle{JHEP}
\bibliography{biblio.bib}

\end{document}